\documentstyle[12pt]{article}
\begin{document}
\baselineskip 7mm
\title{ Measuring the Hausdorff Dimension of Quantum Mechanical Paths }
\author{ H. Kr\"oger$^{a}$, S. Lantagne$^{a}$, 
K.J.M. Moriarty$^{b}$ and B. Plache$^{b}$ \\
$^{a}$ D\'epartement de Physique, Universit\'e Laval, \\
Qu\'ebec, Qu\'ebec G1K 7P4, Canada \\
$^{b}$ Department of Mathematics, Statistics and Computing Science \\
Dalhousie University, Halifax, Nova Scotia B3H 3J5, Canada }
\date{ Universit\'{e} Laval preprint LAVAL-PHY 7/94 \\
October 1994 }
\maketitle
\begin{flushleft}
{\bf Abstract} 
\end{flushleft}
We measure the propagator length in imaginary time quantum mechanics by Monte Carlo simulation on a lattice and extract the Hausdorff dimension $d_{H}$.
We find that all local potentials fall into the same universality class 
giving $d_{H}=2$ like the free motion. A velocity dependent 
action ($S \propto \int dt \mid \vec{v} \mid^{\alpha}$)
in the path integral (e.g. electrons moving in solids, 
or Brueckner's theory of nuclear matter) yields 
$d_{H}=\frac{\alpha }{\alpha - 1}$ if $\alpha > 2$ and $d_{H}=2$ if 
$\alpha \leq 2$. We discuss the relevance of fractal pathes in 
solid state physics and in $QFT$, in particular for the Wilson loop in $QCD$.

\begin{flushleft}
PACS index: 03.65.-w, 05.30.-d
\end{flushleft}
\newpage
When considering the geometry of particle propagation in quantum mechanics 
the suitable language is that of path integrals.
Feynman and Hibbs \cite{kn:Feyn65} have shown
that quantum mechanical pathes can be viewed 
as zig-zag lines, which are no-where differentiable, and which exhibit self-similarity when viewed at different length scales.
Abbot and Wise \cite{kn:Abbo81} have shown that quantum mechanical pathes
of free motion are fractal curves \cite{kn:Mand83} of Hausdorff dimension two. 
In imaginary time (Euclidean) quantum mechanics the 
free motion resembles very much the Brownian motion of a classical particle
as has been observed by Nelson \cite{kn:Nels66}. 
It turns out that the average path 
of a classical particle carrying out a Brownian motion (steps $\Delta x$
and $\Delta t$, such that the diffusion coefficient 
$d = \frac{1}{2} \Delta x^{2} / \Delta t$ is finite)
is also a fractal curve of Hausdorff dimension two 
\cite{kn:Schu81,kn:Itzy89}.
The Euclidean path integral of imaginary time quantum mechanics
corresponds to a Wiener measure \cite{kn:Schu81}.
The case of the  Euclidean free scalar field is discussed 
by Itzykson and Drouffe \cite{kn:Itzy89}.
Quite generally, the measure of Euclidean path integrals
gives the dominant contributions coming from field configurations
corresponding to pathes of no-where differentiable curves \cite{kn:Glim81}.
By measuring the length $L$ of a curve in terms of an elementary length
$\Delta x$, the geometrical property of being a fractal is captured in the Hausdorff dimension $d_{H}$, defined by
\begin{equation}
L \sim_{\Delta x \rightarrow 0} L_{0} \; \left( \Delta x \right)^{1-d_{H}}.
\end{equation}
Strictly speaking, this gives a fractal dimension $d_{f}$ which in most cases coincides with the Hausdorff dimension $d_{H}$.
In quantum mechanics, $\Delta x \rightarrow 0$ requires 
$\Delta t \rightarrow 0$, one is at the continuum limit.
The Hausdorff dimension is a critical exponent.

\bigskip

In this letter we want to investigate the critical exponent
$d_{H}$ of quantum mechanics in the presence of local potentials
and velocity-dependent potentials in the action of the path integral
via Monte Carlo simulations on a temporal
lattice. In order to be able to do a Monte Carlo simulation we consider
imaginary (Euclidean) quantum mechanics with a negative action
\begin{equation}
S[\vec{x},\epsilon] = - \frac{1}{\hbar} \sum_{j=0}^{N-1} \epsilon \frac{m}{2} 
\left( \frac{ \vec{x}_{j+1} - \vec{x}_{j} }{ \epsilon } \right)^{2},
\end{equation}
where we have discretized the time $T= t_{f}-t_{i}$, 
$\Delta t = T/N \equiv \epsilon$. We define the quantum mechanical expectation value of the length for propagating the particle from  $(\vec{x}_{i},t_{i})$
to $(\vec{x}_{f},t_{f})$ in $D$ space dimensions via
\begin{equation}
<L_{f,i}(\epsilon)> = < \sum_{k=0}^{N-1} 
\mid \vec{x}_{k+1} - \vec{x}_{k} \mid  >
= \left. \frac{ \int d^{D}x_{1} \cdots d^{D}x_{N-1} 
\sum_{k=0}^{N-1} \mid \vec{x}_{k+1} - \vec{x}_{k} \mid
\exp[S[\vec{x},\epsilon]] }
{ \int d^{D}x_{1} \cdots d^{D}x_{N-1} \exp[S[\vec{x},\epsilon]] }             
\right|_{\vec{x}_{0}=\vec{x}_{i},\vec{x}_{N}=\vec{x}_{f}}.
\end{equation}
It is useful to introduce another quantity related to the length squared
\begin{equation}
<SQ_{f,i}(\epsilon)> = 
< \sum_{k=0}^{N-1} ( \vec{x}_{k+1} - \vec{x}_{k} )^{2} >.
\end{equation}
This can be computed analytically
\begin{equation}
<SQ_{f,i}(\epsilon)> = \frac{D \hbar \epsilon}{ 2 m }
\left[ \frac{2 m}{ D \hbar} \frac{ ( \vec{x}_{f} - \vec{x}_{i} )^{2} } 
{ T } -2 \right]
+ \frac{ D \hbar T } { m }.
\end{equation}
In the continuum limit ($\Delta t \rightarrow 0$,
$ \mid \Delta \vec{x} \mid \rightarrow 0$), we expect 
critical behavior
\begin{eqnarray}
<L_{f,i}> & \sim & ( \Delta t )^{-A} 
\sim ( \Delta x )^{1-d_{H}},
\\
<SQ_{f,i}> & \sim & ( \Delta t )^{-B} 
\sim ( \Delta x )^{C}.
\end{eqnarray}
In order to determine the Hausdorff dimension, one needs to know 
the elementary length $\Delta x$. 
From the action one can estimate this in $D=1$ dimension from
$\frac{ \Delta t }{\hbar} \frac{m}{2} 
( \Delta x / \Delta t )^{2} \approx 1$,
to give in $D$ dimensions
$< \mid \Delta \vec{x} \mid >_{est} \approx 
\sqrt{ 2 D \hbar \Delta t / m  }$.
From that we can estimate $< L_{f,i} >$ and $< SQ_{f,i} >$
\begin{eqnarray}
< L_{f,i}(\epsilon) >_{est} &=& N < \mid \Delta \vec{x} \mid > 
= T \sqrt{2 D \hbar / m \Delta t }
= 2 D \hbar T / m  
\left( < \mid \Delta \vec{x} \mid > \right)^{-1},
\\
< SQ_{f,i}(\epsilon) >_{est} &=& N < ( \Delta \vec{x} )^{2}>
= 2 D \hbar T / m .
\end{eqnarray}
Comparing eq.(8) with eq.(6) yields $A=1/2$ and $d_{H}=2$ 
confirming the known result for free motion.
Eq.(9) is in agreement (apart from an overall factor 2)
with the exact result of eq.(5) 
(for $\epsilon \rightarrow 0$) and yields $B=C=0$. 

\bigskip

Results of $< L_{f,i} >$ and $< SQ_{f,i} >$
from Monte Carlo simulation are shown in Fig.[1].
We have considered in $D=3$ space dimensions the free motion as well as 
motion in the presence of a local potential $V(\vec{x})$, 
i.e. the cases: (a) $V=0$ (free motion), (b) $V= V_{0} \vec{x}^{2}$
(harmonic oscillator), (c) $V = V_{0} / \mid \vec{x} \mid$ 
(Coulomb potential). We have plotted $\log <L>$ versus $\log N$
and $<SQ>$ versus $N$, where $T = N \Delta t$ is kept fixed.
The expectation value $< \mid \Delta \vec{x} \mid >$ 
is obtained by $<L>/N$. In Fig.[1a,b] we show $<L>$ and $<SQ>$
for the free case and find agreement with the
estimated scaling behavior given by eq.(6-9). In Fig.[1c] we display 
$<L>$ for the harmonic oscillator.
Within numerical errors we find for the free motion as well as for both 
local potentials the same slope.
We have extracted the Hausdorff dimension $d_{H}$ 
which in all cases is found to be consistent within numerical errors
with $d_{H}=2$. Thus in the language of critical phenomena 
these local potentials fall all into the same universality class.
For the free motion $<SQ>$ is consistent with
the exact result from eq.(5). Again the presence of the local potentials does not change the critical exponent $C$ which is compatible with zero.

\bigskip

The above estimate of the Hausdorff dimension hinges upon the 
presence of the velocity squared term in the action.
Thus one expects a different fractal dimension if a term with 
a different velocity dependence is present in the action of the path integral.
Such a velocity dependence occurs in solid state physics 
for electrons moving in solids
\cite{kn:Made81}. Another example in nuclear physics is
Brueckner's theory of nuclear matter \cite{kn:Brue55}, using a 
potential of the form
$V(k) = V_{0} + V_{2} k^{2} + V_{4} k^{4}$.
One should note, however, that a non-Gaussian momentum dependence in the 
quantum Hamiltonian does {\em not} correspond to the same functional form in the action of the path integral. Assuming a polynomial momentum dependence, e.g. $p^{4}$,
in the Hamiltonian leads in the action to a term which falls off faster than a Gaussian as a function of $\Delta x / \Delta t^{1/4}$.
This can be parametrized by 
$ \exp [ - (\Delta x / \Delta t^{1/4} )^{A_{1}} + \cdots 
+ (\Delta x / \Delta t^{1/4} )^{A_{M}} ] $. 
If $A=4/3$, this leads to a velocity dependence $\int dt v^{4/3}$ in the action.
All terms in the above parametrization lead to a $\Delta x \sim \Delta t$
behavior different from that of the free motion 
$\Delta x \sim \sqrt{\Delta t}$. 
In order to keep matters simple, we have chosen to consider
the following parametrization of the action of the path integral
\begin{equation}
S= - \int dt \; \frac{1}{2} m \vec{v}^{2} + V_{0} \mid \vec{v} \mid^{\alpha}
+ U(\vec{x}).
\end{equation}
Here $U$ denotes a local potential. We have seen above that a local potential does not change the fractal dimension, so we drop the $U$ term in what follows.
Let us estimate the Hausdorff dimension for the action
$S= - \int dt V(\vec{v})$.
In analogy to the free case we have
$\Delta t V_{0} 
( < \mid \Delta \vec{x} \mid > / \Delta t )^{\alpha}
\propto 1$,
and thus
\begin{equation}
< \mid \Delta \vec{x} \mid > \; \propto \; 
\left( \Delta t \right)^{\frac{ \alpha -1} { \alpha } }.
\end{equation}
This implies 
\begin{eqnarray}
< L_{f,i} (\epsilon) >_{est} 
&\propto&
\left( \Delta t \right)^{- \frac{1}{\alpha}}
\; \propto \; 
< \mid \Delta \vec{x} \mid >^{ - \frac{ 1 } { \alpha -1 } }_{est},
\\
< SQ_{f,i} (\epsilon) >_{est} 
&\propto&
\left( \Delta t \right)^{\frac{ \alpha -2}{\alpha}}
\; \propto \;
< \mid \Delta \vec{x} \mid >^{ \frac{ \alpha -2 } { \alpha -1 } }_{est},
\end{eqnarray}
Of course, for $\alpha=2$ the scaling behavior agrees with that of the free motion, eqs.(8,9).
From eq.(12), we imply for the Hausdorff dimension 
$d_{H} = \frac{ \alpha } { \alpha -1 }$ and for the critical exponent 
$C=\frac{\alpha -2}{\alpha -1}$.
This opens the mathematical possibility
of having a very large Hausdorff dimension, if $\alpha$ is above but close 
to unity. However, physically this is not realized,
because the presence of the kinetic term in the action
limits $d_{H} \leq 2$,
as will be shown below.

\bigskip

The results of the Monte Carlo simulations with action given by eq.(10) corresponding to $D=3$ dimensions are shown in Figs.[2,3] 
In Fig.[2] we display $<L>$ and $< SQ >$ versus $\log N$ for $\alpha = 1/2$.
We have extracted the critical exponents $d_{H}$ and $C$ as a function of $\alpha$ and show a plot in Fig.[3]. 
We find within numerical errors the following behavior:
For $\alpha < 2$, $d_{H}$ agrees with the value of the free motion, i.e., 
$d_{H}=2$, while for $\alpha > 2$, $d_{H}$ agrees with the value 
$\frac{\alpha}{\alpha -1}$, i.e., the value expected from 
the velocity dependent potential without kinetic energy.
Thus $\alpha_{crit}=2$ is a critical value
of the velocity dependent potential. The same kind of behavior is seen for the 
critical exponent $C$.
  
\bigskip

To conclude let us discuss some perspectives for solid state physics and 
for $QCD$. Consider in solid state physics a semiconductor 
where conduction properties (due to doping) change with position,
e.g., like in a transitor. The Hausdorff dimension is sensitive to 
higher order terms in the energy-momentum dispersion relation.
Although the determination of $d_{H}$ needs a non-local observable,
i.e., $<L>$, the absolute length is not 
relevant for $d_{H}$. So in principle one could measure $<L>$ 
on a length scale large to atomic distance, but small to 
the length of variation of macroscopic properties.
In such a way, one could determine $d_{H}$ locally and study its 
change as a function of position throughout the semiconductor
(this would correspond to a velocity dependent potential, where 
$\alpha$ is a function of position). In this sense $d_{H}$ 
can be viewed as one parameter probing the local geometry
of quantum mechanics.
In $QCD$, a fractal dimension might eventually play a role in a non-local observable, like the Wilson-loop. The Wilson loop distinguishes between 
a confined (area law) and a non-confined (perimeter law) phase. It also has a quark-quark potential interpretation which yields the string tension, being crucial for the baryon spectrum. The potential interpretation makes the semiclassical assumption that the quarks propagate along classical 
smooth curves. In quantum mechanics however, one would expect the quarks to propagate along zig-zag lines. The path of quark propagation 
enters in the definition of the Wilson loop.
Going over from smooth curves, $d_{H}=1$, to zig-zag lines,
e.g., with $d_{H}=2$ which are similar to a surface, would most likely change the behavior of this observable.

\bigskip

{\bf Acknowledgement}
H.K. is grateful to J. Polonyi and 
V. Branchina for helpful discussions.
H.K. was supported by NSERC Canada and FCAR Qu\'ebec.

\newpage
\begin{flushleft}
{\bf Figure Caption}
\end{flushleft}
\begin{description}
\item[{Fig.1}]
(a) Length $<L>$, and (b) sum of length squares $< SQ>$ for free motion
and (c) length $<L>$ for harmonic oscillator versus number of time steps $N$.
\item[{Fig.2}]
Same as Fig.[1] for velocity dependent action ($\alpha =1/2$). 
\item[{Fig.3}]
Critical exponents $d_{H}$ and $C$ versus potential parameter $\alpha$
for velocity dependent action. The full line gives the estimated exponents
from eq.(11-13).
\end{description}
\end{document}